\documentstyle[sprocl_mod,psfig,epsfig]{article}
\newcommand{\mz}{M_{_Z}}
\newcommand{\afba}[1]{A^{#1}_{_{\rm FB}}}
\begin{document}
\thispagestyle{empty}
\begin{flushleft}
{\tt
DESY 99-166
\\
November 1999
\\
hep-ph/9911296
}
\end{flushleft}

\bigskip

\title{%
QED RADIATIVE CORRECTIONS TO $e^+e^-\to \bar{f} f$\\
WITH REALISTIC CUTS\\
AT LEP ENERGIES AND BEYOND
\footnote{Talk presented at
14th International Workshop on High Energy Physics and Quantum Field Theory
(QFTHEP~'99), Moscow, Russia, May 27 - June 2, 1999; to appear in 
the Proceedings.
}}
\author{%
M. JACK
}
\address{%
DESY Zeuthen, Platanenallee 6,
\\ 
D-15738 Zeuthen, Germany
\\ 
E-mail: jack@ifh.de}
%

\maketitle

\abstracts{ 
After 10 years of steadily increasing the experimental precision  
at LEP/SLC, there is a strong demand on an update of existing
programs for fermion pair production.
We present a rederivation of the $O(\alpha)$
Bremsstrahlung corrections to $e^+e^-\to\bar{f}f$ for
the semi-analytic program {\tt ZFITTER}. We focus on observables 
like total cross section and forward-backward asymmetry
in the leptonic case with combined cuts on acollinearity angle, 
acceptance angle, and minimal energy of the fermions.
The outcome of our analysis is a shift of the predictions
by {\tt ZFITTER} at LEP 1 energies off-resonance of a few per mil
while at the $Z$ resonance numerical changes can be neglected.
Thus we obtain for cross sections and asymmetries at LEP 1  
a level of agreement with other programs of better than per mil, 
like for the kinematically simpler $s'$ cut option.
  A preliminary analysis of {\tt ZFITTER}, {\tt TOPAZ0}, 
and other codes at LEP 2 energies showing deviations 
of several per cent with acollinearity cuts 
enforce a future examination of higher order effects 
with different cuts.
The predictions by LEP/SLC data, however, are not affected
within the experimental errors. 
}


\section{\Large \bf Introduction}
\label{sec_intro}

In the light of high precision measurements to the Standard Model 
({\tt SM}), fermion pair production in $e^+e^-$ annihilation still plays a
very important role, e.g. for the extraction of information from data 
on the electroweak symmetry breaking sector of the {\tt SM}.
This impressing increase of precision by experiment over the last 10
years was nicely illustrated by \cite{Swartz:1999lp} : 
%
\begin{table}[htb]
\begin{center}
\renewcommand{\arraystretch}{1.1}
\begin{tabular}{|c||c|c|}
\hline
Quantity & LP~89 (233 events) & LP~99 
($18\times 10^6$ events)
\\
\hline
$M_Z$ (GeV)  &  $91.17\pm 0.18$  &  $91.1871\pm 0.0021$
\\
$\Gamma_Z$ (GeV)  &  $1.95^{+0.40}_{-0.30}$  &  $2.4944\pm 0.0024$
\\
$N_\nu$ (light) &  $3.0 \pm 0.9$  &  $2.9835 \pm 0.0083$
\\ 
\hline 
\end{tabular}
\caption[]{
{\sf
Examples for the development of high 
precision measurements at the $Z$ resonance.}
}
\label{Zprecision}
\end{center}
\end{table}

As one of the most 
copious processes from LEP to Linear Collider (LC) energies
fermion pair production also forms an important incoherent 
background to W-pair production, or more generally, to 4-fermion 
final states at higher energies \cite{Boudjema:1996qg,Accomando:1997wt}.
Another very interesting application of this `traditional' channel 
is the possibility to extract limits on extensions of the {\tt SM} 
in experiments like LEP or SLC when comparing with the theoretical 
predictions (e.g. $Z'$ and $W'$ boson searches, contact interaction 
scales, exchange of leptoquarks, preons, or particles in R-parity 
breaking supersymmetric models \cite{Christova:1998tc}).

We want to focus on high precision measurements to the {\tt SM} and
{\tt MSSM}. The needed accuracy at energies below the $Z$ resonance region 
from QED Bremsstrahlung, forming the bulk of the radiative corrections,
is relatively small. However, with the high experimental precision now  
obtained at LEP~1 and LEP 2 energies, it is necessary to estimate
theoretically these corrections around and above the $Z$ resonance
region. Especially, hard QED corrections which become more and more 
important with growing c.m. energies and resummed soft and virtual 
higher order effects have to be calculated precisely taking into
account realistic experimental cuts.  
Having this in mind, the accuracy of the theoretical prediction 
for single contributions to observables should be typically a factor 
of 10 or so better than the experimental error.
This is moreover true when considering a high luminosity option
of a future LC running at the $Z$ resonance
(`Giga $Z$' option) \cite{Moenig:1999aa}. 
Looking at the experimental situation (Table \ref{Zprecision}), 
we can briefly state that the theoretical predictions therefore 
have to be better than 0.015\% 
at the peak, at the order of 0.03\% 
in the resonance region ($\approx M_Z\pm 3\,\mbox{GeV}$),
and around 0.5\% 
starting at LEP 2 and higher energies.

\section{\Large \bf Realistic observables and the {\tt ZFITTER} concept}
\label{sec_conc}

The semi-analytic approach of the {\tt ZFITTER} code \cite{DESY99070} 
consists of a fast, one-dimensional numerical integration
of analytical formulae for different observables like cross sections, 
asymmetries, polarizations, and angular distributions with the inclusion 
of different experimentally relevant cut options and resummed  
$O(\alpha)$ QED corrections plus dominant higher order effects. 
The number of analytically calculable cuts in such an aproach is of
course limited, in our case to two angular cuts and to cuts
on the final state energies and invariant mass squared.
Corresponding numerical programs for fermion pair production 
like {\tt ALIBABA} \cite{Beenakker:1991mb}, {\tt BHM} \cite{Burgers:BHM},
{\tt KORALZ, KK} \cite{Jadach:1994yv,Jadach:1999kz},
or {\tt TOPAZ0} \cite{Montagna:1998kp}
are in this respect complementary to our approach, as they can in principle  
treat multi-differential observables with nearly 
arbitrary cuts to the final state phase space, 
but this at the expense of a clear
increase in computing time.

In the {\tt ZFITTER} approach
\cite{Bardin:1989di,Bardin:1989cw,Bardin:1991de,Bardin:1991fu,DESY99070},
we calculate:
\begin{eqnarray}
\sigma(s) \sim \int \frac{ds'}{s}~ {\sigma^0(s')}~ \rho(s'/s) ,
\label{siggen}
\end{eqnarray}
with $s'=m^2_{f\bar f}$ as the invariant mass squared of the 
final state fermion pair.
Cross sections and asymmetries like $\sigma_T(s)$ and $A_{FB}(s)$
are treated in an {\em improved Born approximation}, convoluting   
{\em effective Born observables}, $\sigma^0(s)$ and $A^0(s)$, over $s'$
with a flux function $\rho$ ({\em radiator}) containing the photonic 
corrections. The electroweak and QCD corrections are described
with effective couplings in the effective Born terms; 
this is valid due to their smallness at LEP 1 energies.~\footnote{
At higher energies, $ZZ$ and $WW$ box corrections
contribute at the 1 to 2 per cent level at LEP 2 energies so 
the validity of our non-gauge invariant, effective Born 
approximation will have to be carefully reexamined there. 
}
With {\tt ZFITTER}, three different cut options are available:
(i) no cut \cite{Bardin:1989cw},
(ii) cuts on $s'$ and on the scattering angle $\vartheta$ of one
fermion \cite{Bardin:1991de,Bardin:1991fu}, 
or  
(iii) cuts on the fermions' acollinearity angle, $\theta_{acol}$, 
on their energies, $E^{f}=E^{\bar f}$, and on $\cos \vartheta$
\cite{Christova:1999cc,DESY99037}.
{For} a detailed description of the phase space with cuts on final state 
acollinearity, minimal energies, and/or $s'$ please refer to
\cite{Passarino:1982zp,Christova:1998tc}.
The effective Born cross sections, {$\sigma^0(s')$}, may also be chosen
according to following approaches:
(A) Standard Model,
(B) Model Independent,
(C) Others \cite{Bardin:1989di,Bardin:1992jc2,DESY99070}.
{For} the latest updates and afs-accounts for the {\tt ZFITTER} code 
and other programs please consult 
\cite{DESY99070,Jadach:1999kz,Montagna:1998kp,SITGES99}.

\section{\Large \bf Different codes -- comparison and problems}
\label{sec_codes}

Focussing first on LEP 1 energies and the $s'$ cut branch  
(see Section \ref{sec_conc}), the situation of the {\tt ZFITTER} 
code up to versions v.5.x (1998) in comparison with the code 
{\tt TOPAZ0} can be stated as quite satisfactory. 
At $M_Z\pm 3\,\mbox{GeV}$ the agreement is better than $10^{-4}$ 
\cite{Bardin:1995aa,Bardin:1999gt}. 
And also at LEP 2 energies and higher we can meet the demands 
by experiment with a deviation of the codes of not more than 1 
or 2 per mil for different cut values, and with a substantial 
decrease of this difference below 1 per mil in case of
a sufficiently large $s'$ cut.
This situation does not change when an extra cut on the maximal 
scattering angle $\cos\vartheta$ is applied \cite{DESY99037,SITGES99}. 

If we introduced an acollinearity cut instead of the 
$s'$ cut, we obtained a comparable agreement around the $Z$ resonance 
as long as we only considered initial state Bremsstrahlung. But as 
soon as we included the initial-final state corrections this agreement 
deteriorated to $O(3\times 10^{-3})$ which grows to an 
unbearably large discrepancy between the two codes of several 
per cent at larger energies ($\sqrt{s}\approx 100\ldots
200\,\mbox{GeV}$) \cite{Christova:1998tc}.
An earlier comparison of the {\tt ZFITTER} code with the {\tt ALIBABA} code 
for the s-channel part of the Bhabha scattering 
branch had already shown similar deviations \cite{Riemann:199200,DESY99037}. 

\section{\Large \bf QED Corrections with Acollinearity Cut
\label{sec_photcorr}   }

The formulae for photonic $O(\alpha)$ corrections in {\tt ZFITTER} 
with $s'$ cut have been analytically and numerically multiply 
checked at LEP energies 
\cite{Bardin:1989cw,Bardin:1991de,Bardin:1991fu,Boudjema:1996qg,Bardin:1999gt}.
{For} the acollinearity cut branch, however, 
there has been no independent check until recently and only little
literature available on the exact $O(\alpha)$ final state 
corrections to the total cross section
and forward-backward asymmetry 
\cite{Montagna:1993mf} and on some formulae related to the 
initial state corrections (and its combined exponentiation 
with final state radiation) for the angular distribution 
\cite{Bilenkii:1989zg,MBilenky:1989ab}.
So, a recalculation and documentation of the 
acollinearity cut situation was absolutely mandatory with 
the main focus first at energies around the $Z$ resonance. The 
slightly more involved Bhabha scattering case with extra 
$t$ channel contributions is kept for a later analysis. 

In our semi-analytic approach it is technically feasible to 
calculate observables with cuts on one of the final state fermions' 
scattering angles, $\vartheta$, their energies, $E_{{\bar f},f}$, 
their acollinearity angle, $\theta_{acol}$, and/or their 
invariant mass squared, $s'$.   
The perturbative calculation of the $O(\alpha)$ hard corrections 
to total (differential) cross 
sections then consists of an analytical integration over 3, or 
respectively 2, angles of phase space. The corresponding formulae 
are finally numerically integrated over $s'$. 

We observe that neglecting the initial and final state masses 
at the mentioned high energies necessitates a separation of the 
phase space formed by the remaining two angles of integration, 
$\cos\vartheta$ and $\cos\theta_{acol}$, into several different regions. 
Only where necessary, the masses are kept in order to regularize 
the mass singularities from collinear radiation of Bremsstrahlung
photons. This `slicing' of phase space 
delivers for each region different analytical
expressions for the calculated observables \cite{DESY99037}.
{For} the special cases of either full angular 
acceptance, i.e. no cut on $\cos\theta$, or no cut 
on the acollinearity angle $\theta_{acol}$, the number of different
expressions can be substantially reduced and very compact formulae 
can be obtained \cite{Christova:1999cc}. 
This allows one to install a numerically fast second branch in the Fortran 
code when demanding fewer cuts, important for quick data-fitting 
routines needed by experiment.

The Fortran package {\tt acol.f} contains a complete collection of the 
analytical formulae for the $O(\alpha)$ corrections and is called 
from {\tt ZFITTER} v.6.04/06 \cite{zfitter:v6.0406m} onwards. 
The angular distribution will be available in versions v.6.2x. 
A complete collection of all analytical expressions is in preparation
\cite{DESY:1999??}.

\section{\Large \bf Results}
\label{sec_res}

The main modifications in the new coding are corrected terms 
in the QED initial state and interference radiator parts. For 
different regions of phase space, these are different, 
non-logarithmic contributions proportional to $\cos\vartheta$
in the symmetric part of the angular distribution 
$d{\sigma}/d{\cos\vartheta}$ and terms of the type 
$a+b\cos^2\vartheta$ in the antisymmetric part with corresponding
changes in the integrated results \cite{DESY99037,DESY:1999??}. 
Omitting these terms in earlier versions is justified with the then 
anticipated experimental precision of only $5\times 10^{-3}$ at LEP 1,
but not with the higher accuracy now at the $Z$ resonance peak.

The net effect of the corrected initial state, final state, and
interference terms is depicted for $\sigma_T$ and $A_{FB}$ in
Fig. \ref{netsigafb} for the energy range $30$ to $300\,\mbox{GeV}$
(default flag setting):  
%
\begin{figure}[htb] 
\begin{flushleft}
\begin{tabular}{ll}
  \mbox{%
\epsfig{file=%
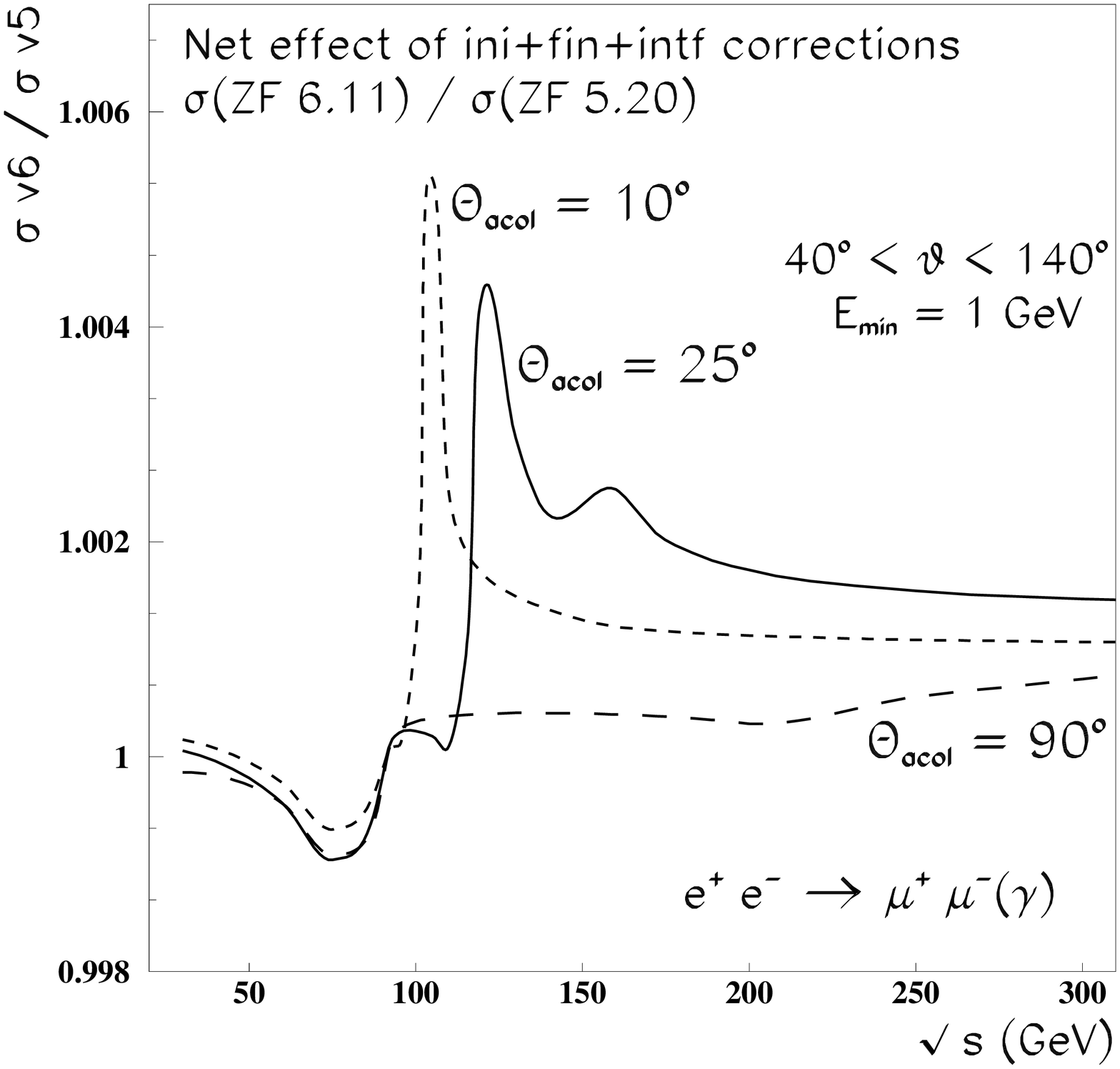
           ,width=7.7cm   
         }}%
&
  \mbox{%
\epsfig{file=%
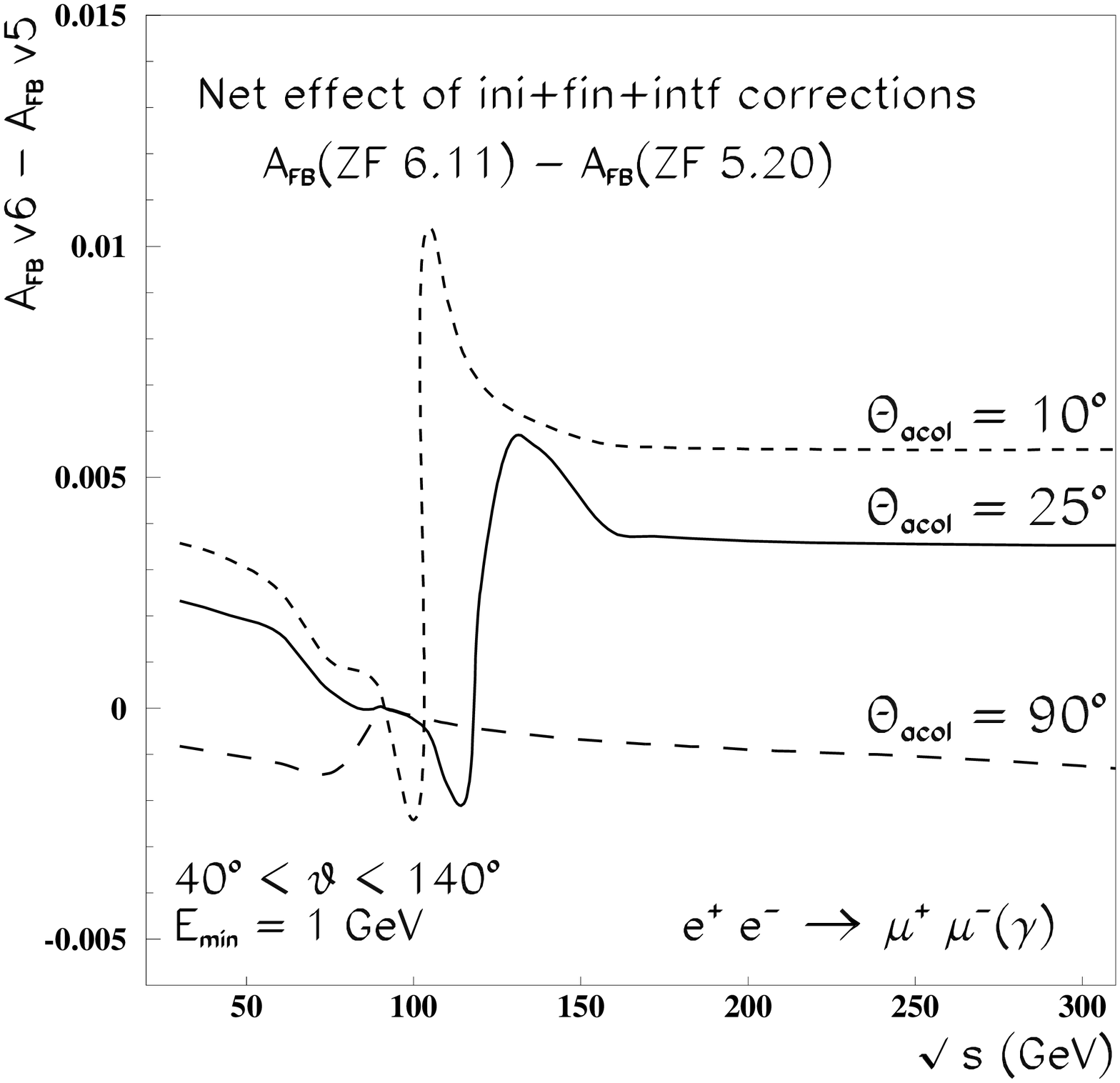
           ,width=7.7cm   
         }}%
\\
\end{tabular}
\caption
{\sf
Net ratios of muon pair production cross sections and differences of 
forward-backward asymmetries predicted from {\tt ZFITTER} 
v.6.11 and v.5.20 
with three different
acollinearity cuts: $\theta_{\rm acol} < 10^{\circ}, 25^{\circ}, 90^{\circ}$;
$E_{min}=1\,\mbox{GeV}$; $40^\circ<\vartheta<140^\circ$; 
all corrections included.
\label{netsigafb}
}
\end{flushleft}
\end{figure}
%
We compared the cross section ratios and the absolute 
differences of the asymmetries of {\tt ZFITTER} v.6.11
\cite{DESY99070}, containing the 
new results, with v.5.20 \cite{zfitter:v5.2021m}, 
still with the old coding, for 
different acollinearity cut values.
The net corrections are largest where the radiative 
return to the $Z$ starts to be prevented by the acollinearity 
cut. {For} $\theta_{acol} < 10^\circ$ or $25^\circ$ this sets in 
at roughly $\sqrt{s} > 100\,\mbox{GeV}$, or $115\,\mbox{GeV}$
respectively. These corrections to the code are at most roughly 0.5\%
for $\sigma_T$ and 1\% 
for $A_{FB}$ and shrink below 1\% at higher energies.
   
The corrections are mainly due to the new initial-final state interference 
contributions. This is illustrated for a maximal acollinearity 
angle of $\theta_{acol} < 10^\circ$ in Table \ref{tab-acol10-th40} 
where the shifts of the initial state 
corrected total cross sections and asymmetries are shown 
when switching on the interference contributions. 
One can show that the numerical effects stay below 1\% 
at LEP 2 energies with such a cut value
\cite{DESY99037,SITGES99}. 
At LEP 1 they are much smaller, of the order of few per mil.
%
\begin{table}[htb]
\begin{center}
\renewcommand{\arraystretch}{1.1}
\begin{tabular}{|c||c|c|c|c|c|}
\hline
\multicolumn{6}{|c|}{
{$\sigma_{\mu}\,$[nb] with $\theta_{\rm acol}<10^\circ$}}
\\ 
\hline
$\theta_{\rm acc} = 0^\circ$& $\mz - 3$ & $\mz - 1.8$ & $\mz$ & $\mz + 1.8$ &
$\mz + 3$  \\ 
\hline\hline
  & 0.21932  & 0.46287  & 1.44795  & 0.67725  & 0.39366 \\
{{\tt TOPAZ0}}  & 0.21776  & 0.46083  & 1.44785  & 0.67894  & 0.39491 \\
  & 
{\bf --7.16}    & 
{\bf --4.43}     &
{\bf --0.07}     &
{\bf +2.49}     &
{\bf +3.17}    
\\ 
\hline
  & 0.21928  & 0.46284  & 1.44780  & 0.67721  & 0.39360 \\
{{\tt ZFITTER}}  & 0.21772  & 0.46082  & 1.44776  & 0.67898  &
0.39489 \\ 
  &
{\bf --7.16}     &{\bf --4.40}     &{\bf --0.03 }    &{\bf +2.60}
&{\bf +3.27}   
\\ 
\hline 
\hline
\multicolumn{6}{|c|}{$\afba{\mu}$ with $\theta_{\rm acol}<10^\circ$} \\
\hline
$\theta_{\rm acc} = 0^\circ $& $\mz - 3$ & $\mz - 1.8$ &
$\mz$ & $\mz + 1.8$ & $\mz + 3$  \\ 
\hline\hline
  & --0.28450 & --0.16914  & 0.00033  & 0.11512  & 0.16107 \\
{{\tt TOPAZ0}}  & --0.28158 & --0.16665  & 0.00088  & 0.11385  &
0.15936 \\ 
  & 
{\bf +2.92}    & {\bf +2.49}     &
{{\bf +0.55}}     &{\bf --1.27}
&
{{\bf --1.71}}    \\ 
\hline
  & --0.28497 & --0.16936  & 0.00024  & 0.11496  & 0.16083 \\
{{\tt ZFITTER}}  & --0.28222 & --0.16710  & 0.00083  & 0.11392  & 0.15926 \\
  & {\bf +2.75}    & {\bf +2.27}     & {\bf +0.60}    &{\bf --1.03}
&{\bf --1.56}
\\
\hline 
\end{tabular}
\caption[]{
{\sf
A comparison of predictions from {\tt ZFITTER} v.6.11 and {\tt TOPAZ0}
v.4.4 for muonic cross sections and forward-backward
asymmetries around the $Z$ peak.
First row is without initial-final state interference, second row with,
third row the relative effect of that interference in per mil.
}
}
\label{tab-acol10-th40}
\end{center}
\end{table}

The corrected initial state and final state terms
only have minor effects on $\sigma_T$ and $A_{FB}$ 
and amount at most to corrections at the order of 
0.1\% -- 0.2\% for $A_{FB}$ at LEP 2 energies for
different cuts. 
A detailed analysis of all new modifications to the code
can be found in \cite{DESY99037}. 

\begin{figure}[htb] 
\begin{flushleft}
\begin{tabular}{ll}
  \mbox{%
\epsfig{file=%
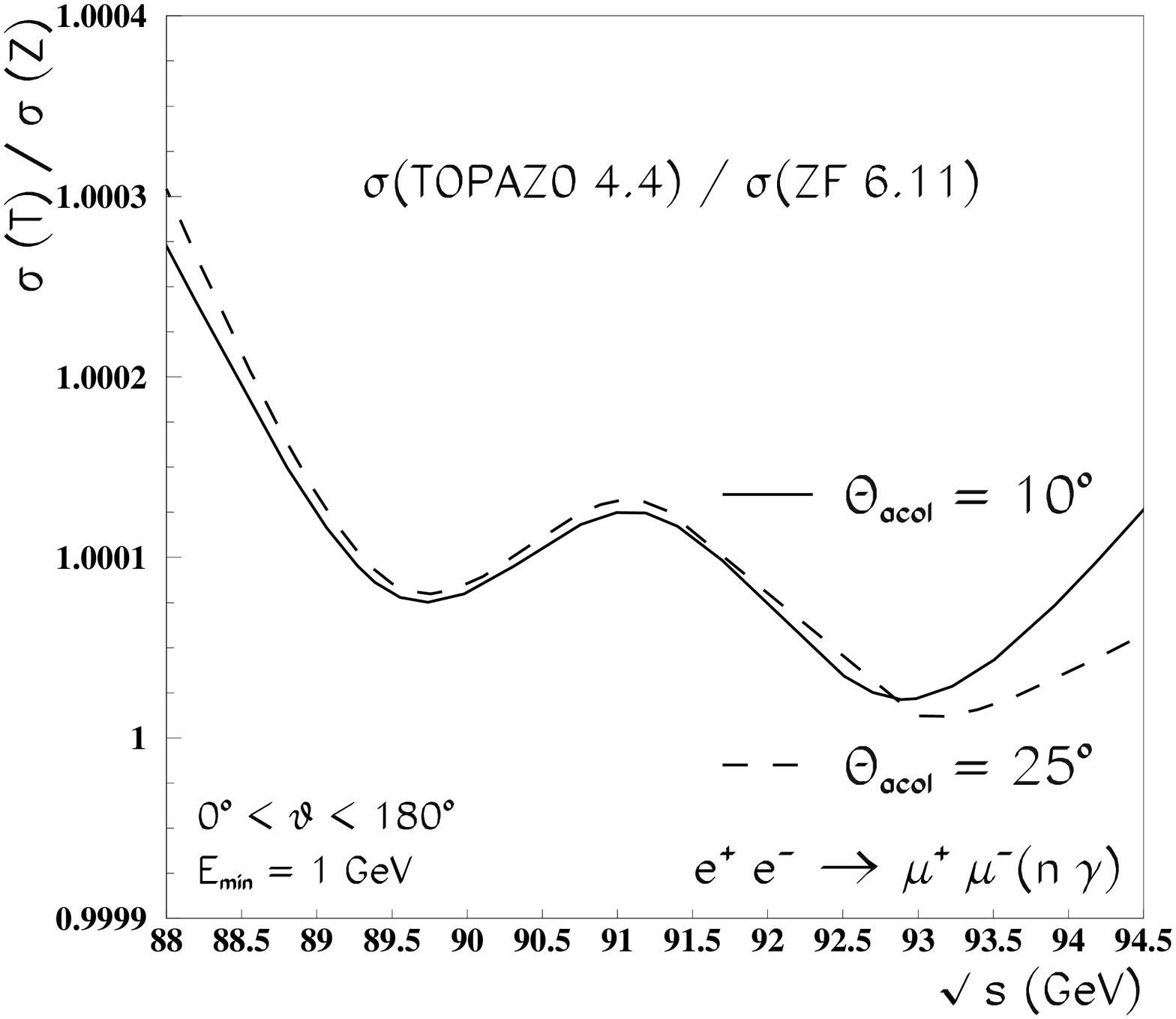
           ,width=7.7cm   
         }}%
&
  \mbox{%
\epsfig{file=%
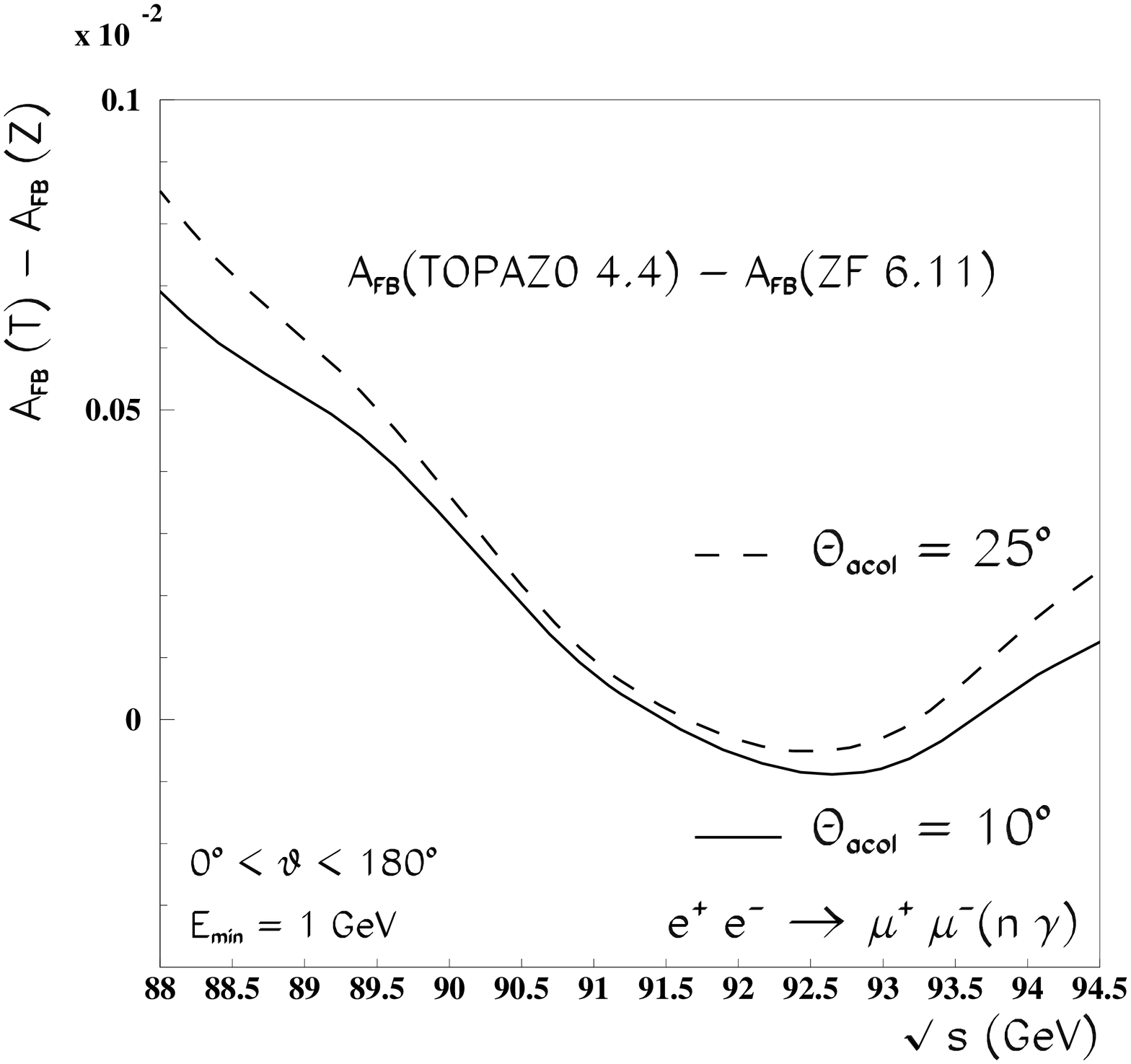
           ,width=7.7cm   
         }}%
\\
\end{tabular}
\vspace{-0.5cm}
\caption
{\sf
Ratios of muon pair production cross sections and differences of 
forward-backward asymmetries at the $Z$ resonance: 
{\tt ZFITTER} v.6.11 vs. {\tt TOPAZ0} v.4.4 
for different acollinearity cuts
(full acceptance).
\label{top-zf-peak}
}
\end{flushleft}
\end{figure}

The numerical comparison of the newly updated {\tt ZFITTER} version 
v.6.11 \cite{DESY99070} 
with {\tt TOPAZ0}'s latest release version 
v.4.4 \cite{Montagna:1998kp} 
now delivers for LEP 1 energies the same high
level of agreement as for the $s'$ cut (Fig. \ref{top-zf-peak}): 
At the peak itself we have a deviation of the codes of 
$O(10^{-4})$ or less for $\sigma_T$ and $A_{FB}$, with 
an acceptable increase to $O(3\times 10^{-4})$ for $\sigma_T$
and a slightly worse value for $A_{FB}$ of $O(7\times 10^{-4})$
with a maximal acollinearity angle of $10^\circ$. 
%

\begin{figure}[htb] 
\begin{flushleft}
\begin{tabular}{ll}
  \mbox{%
  \epsfig{file=%
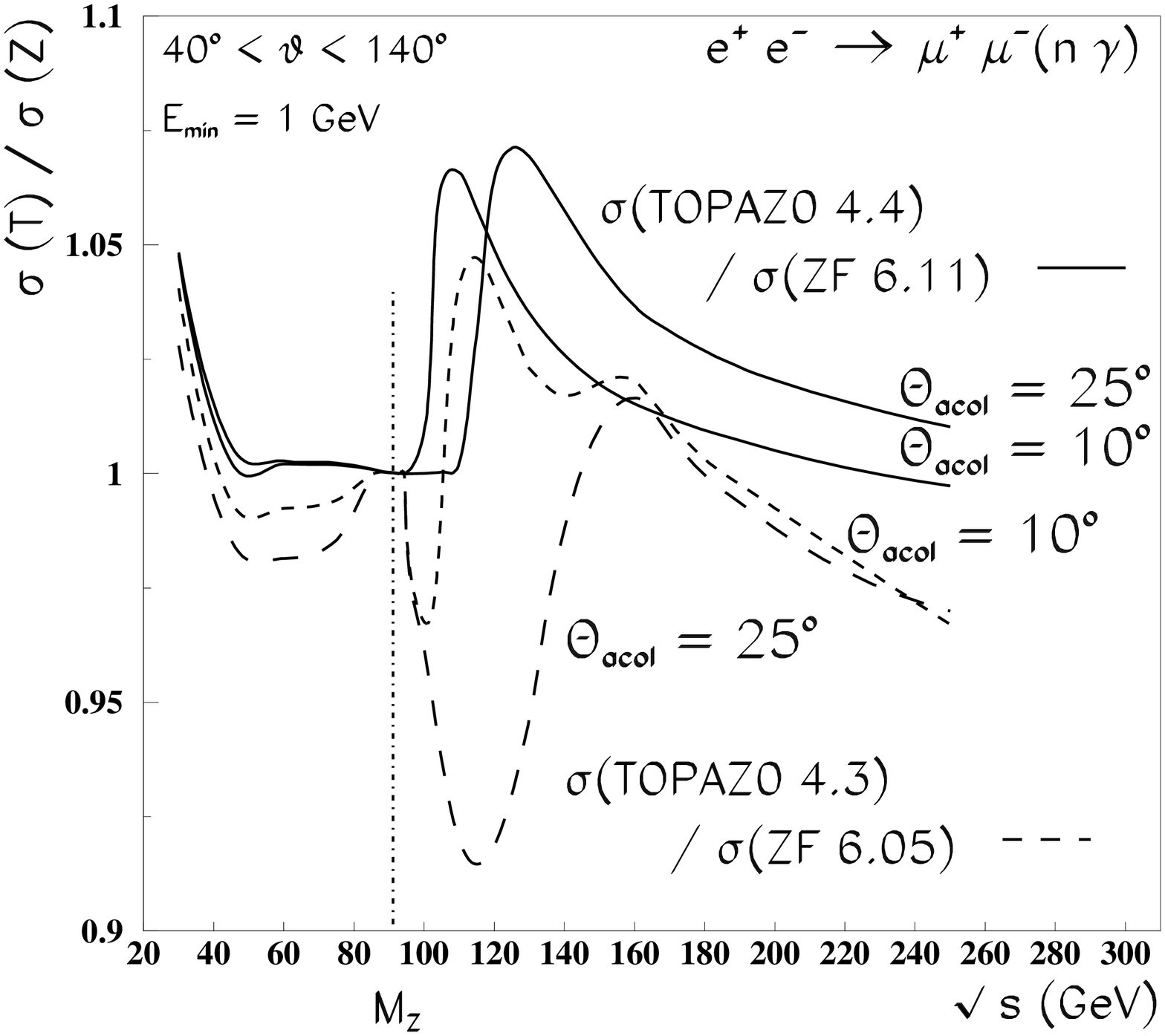
,width=8.cm   
         }}%
&
  \mbox{%
  \epsfig{file=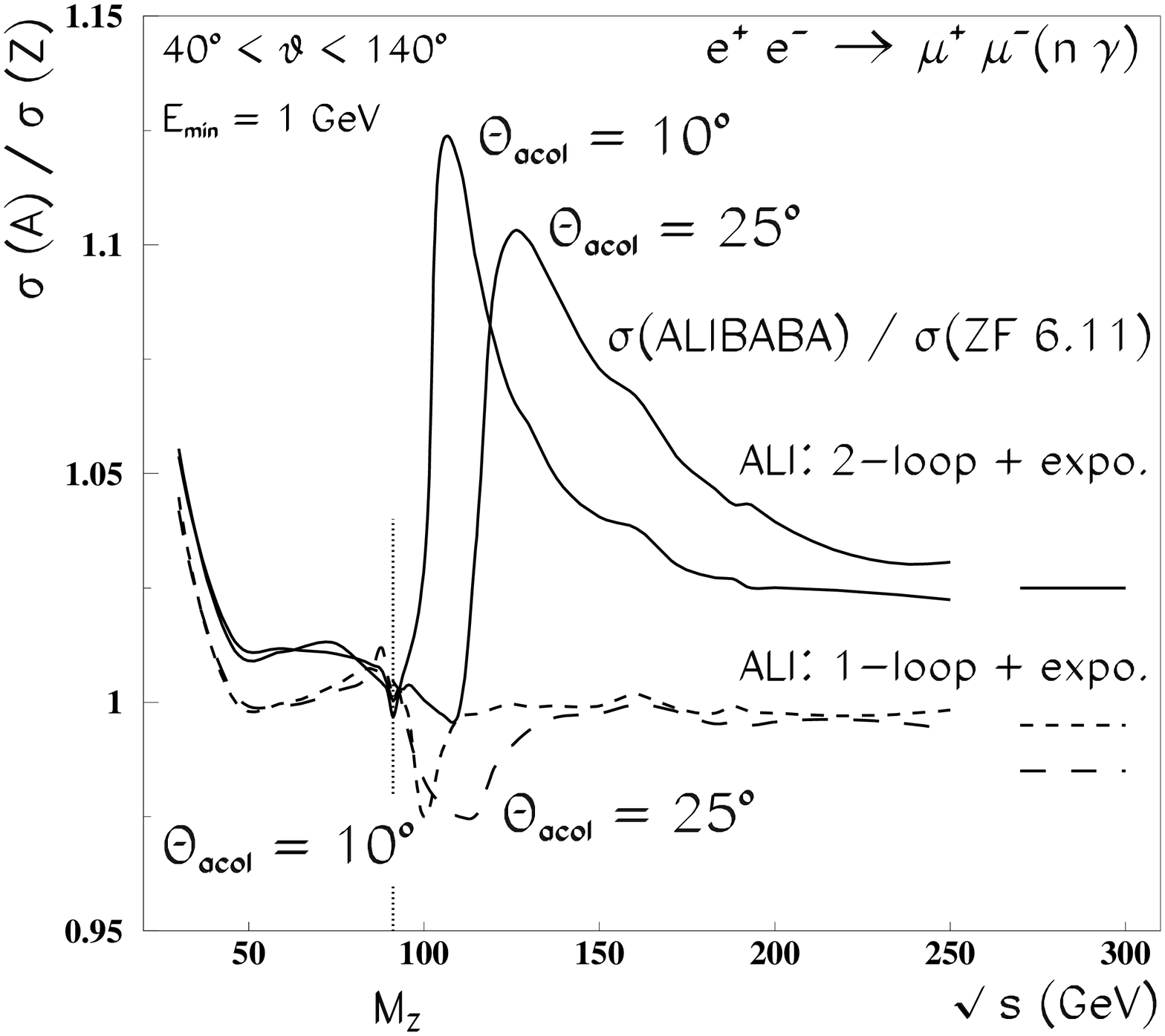,width=8.cm}}
\\
\end{tabular}
\vspace{-0.5cm}
\caption[]
{\sf
Comparison of predictions from {\tt ZFITTER} v.6.04/06 
and v.6.11, {\tt TOPAZ0} v.4.3 and v.4.4, 
and {\tt ALIBABA} v.2 for muon pair 
production cross sections.
\label{ali-top-zf}}
\end{flushleft}
\end{figure}

{For} a wider energy range, we compared     
{\tt ZFITTER} v.6.11 (1999) with {\tt TOPAZ0} v.4.3 (1998) 
and v.4.4 (1999)
\cite{Passarino:199800,Montagna:1998kp,Bardin:1999gt}
and {\tt ALIBABA} v.2 (1990)
\cite{Beenakker:1991mb} in Fig. \ref{ali-top-zf}. 
An older comparison can be found in \cite{Christova:1998tc}.
All numbers have been produced with the default settings of the 
programs.
%

At LEP 2 energies the deviation of 
{\tt ZFITTER} v.6.11 and {\tt TOPAZ0} v.4.4 is at the order of 1\%
or less for different acollinearity cuts and an 
acceptance cut of $40^{\circ} < \vartheta < 140^{\circ}$. 
A cross check with comparable $s'$ cuts delivers a much better
agreement of about 1 per mil or better \cite{DESY99037}.

In both cases, however, there is a clear peak of the cross section ratios 
at energies where the $Z$ radiative return is not prevented by the 
cuts. While for the $s'$ cut this discrepancy stays moderate at the 
per cent level, it grows up to several per cent for the acollinearity 
cut.~\footnote{The flip of sign of these effects compared to the 
older versions, {\tt TOPAZO} v.4.3 and {\tt ZFITTER} v.6.05, is mainly 
due to a corrected interference contribution in the {\tt TOPAZ0} code.
Changes to code {\tt ZFITTER} v.6.05 were negligible here.}  
This effect is now similar to the observed peak in the 
{\tt ALIBABA} - {\tt ZFITTER} comparison
at energies above roughly $100\,\mbox{GeV}$.
Interestingly enough, when switching off two-loop contributions 
present in {\tt ALIBABA}, but not in {\tt ZFITTER}, the agreement 
improved considerably. Corrections by initial state pair production
or different exponentiation of initial and final state higher orders, 
however, do not have a large effect here \cite{SITGES99}. 

Preliminary studies show that a correct description of hard  
two-loop QED corrections, especially for the acollinearity cut 
option in the {\tt ZFITTER} code, together with a correct
resummation of the soft and virtual initial-final state 
interference contribution, not contained in the {\tt ZFITTER} 
code so far, seem to play a key role here.
Since the acollinearity cut is not as effective in preventing the 
radiative return to the $Z$ as the $s'$ cut, these deviations also  
survive more profoundly for the acollinearity cut at higher energies
than for the $s'$ cut.
Further comparisons in this respect for both cut options, especially 
including programs {\tt KORALZ} \cite{Jadach:1994yv} and {\tt KK}
\cite{Jadach:1998jb}, are intended (see also \cite{Jadach:1999gz}).   

\section{\Large \bf Summary}
\label{sec_sum}

We presented a rederivation of analytical formulae for the $O(\alpha)$
hard QED Bremsstrahlung corrections to $e^+e^-\to \bar{f} f$
in the case of leptonic final states with cuts to the 
fermions' acollinearity angle and energies ($f\neq e$).
This was done in the context of the semi-analytic program {\tt ZFITTER} 
calculating radiatively corrected observables with realistic 
experimental cuts, e.g. for LEP/SLC applications. These corrections
to all contributions -- initial state, final state, and initial-final
state interference -- are certain non-logarithmic terms which could be  
neglected earlier but have to be considered now with the new high
level of experimental precision (e.g. $\delta{M_Z}/M_Z 
= 2.2\times 10^{-5}$ \cite{Swartz:1999lp}).

The older versions of {\tt ZFITTER}, i.e. versions v.5.20
\cite{zfitter:v5.2021m} and earlier,
derive the $O(\alpha)$ QED corrections to $\sigma_T$ with 
acollinearity cut with a numerical accuracy of about 
0.4 \% in the $Z$ resonance region ($M_Z\pm 3\,\mbox{GeV}$),
and similarly for $A_{FB}$ with about 0.13 \%. 
The limiting factor here are mainly the initial-final state 
interference corrections. The new and improved coding 
in {\tt ZFITTER} v.6.11 \cite{DESY99070} now reproduces the very nice 
agreement with program {\tt TOPAZ0} v.4.4 \cite{Montagna:1998kp}
already obtained for the $s'$ cut \cite{Bardin:1999gt} 
of better than 0.03\% (0.1\%) for $\sigma_T$ ($A_{FB}$)
at LEP 1 and better than 0.01\% at the $Z$ resonance peak.~\footnote{
At the $Z$ peak itself, the accuracy had already been quite 
satisfactory before, i.e. better than $10^{-4}$, due to  
suppressed hard-photon radiation.
}
At higher energies, $\sqrt{s}> 200\,\mbox{GeV}$, the agreement 
with $s'$ cut is better than per mil, but still only 1 to 2\%
for the acollinearity cut. Especially in the intermediate energy range, 
where the $Z$ radiative return events are not prevented, this discrepancy  
peaks and amounts up to several per cent. A similar effect is also
visible for the $s'$ cut, although to a much lesser extent (below 1\%).

Preliminary studies indicate that higher order corrections,
in particular, hard two-loop corrections together with an
acollinearity cut and a correct resumming of soft and virtual 
interference effects not included in the {\tt ZFITTER} code 
seem to be the major underlying cause for this last point 
\cite{Jadach:1998jb,Jadach:1999gz}. Further studies hereof, also for 
the angular distribution, and a similar update of the code for the 
Bhabha scattering case are in preparation.

\section*{\Large\bf Acknowledgments}
I am indebted to P.~Christova, S.~Riemann, and 
T.~Riemann for their kind support and cooperation
and for many helpful discussions on the presented issues here.
I would also like to thank J.~Illana, T.~Riemann, and A.~Tkabladze 
for carefully reading the manuscript. I finally would very much like 
to thank the Organizing Committee of the QFTHEP~'99 Workshop for their warm 
and kind hospitality throughout the workshop.

\section*{\Large\bf References}


\begingroup\begin{thebibliography}{100}
\bibitem{Swartz:1999lp}
M.~Swartz, ``Precision Electroweak Physics at the Z'', talk at 
LEPTON-PHOTON~'99, Aug 9-14, 1999, Stanford, U.S.A.,
to appear in the proceedings.

\bibitem{Boudjema:1996qg}
E.~Accomando {\em et~al.}, ``Standard model processes'', in {\em Physics at
  {LEP2}, {\rm report CERN 96--01 (1996)}} (G.~Altarelli, T.~Sj{\"o}strand, and
  F.~Zwirner, eds.), pp.~207--248, {\tt hep-ph/9601224}.

\bibitem{Accomando:1997wt}
ECFA/DESY LC Physics Working Grp. Coll., 
E.~Accomando {\em et~al.}, {\em Phys. Rept.} {\bf 299} (1998) 1.

\bibitem{Christova:1998tc}
P.~Christova, M.~Jack, S.~Riemann, and T.~Riemann,  
``Predictions for fermion-pair production at LEP'', 
in {\em Proceedings of Fourth Int. Symposium on Radiative
Corrections, Barcelona, Spain, Sep 8-12, 1998} (J.~Sola ed.),
p.~392, World Scientific, Singapore, 1999. 

\bibitem{Moenig:1999aa}
K.~M{\"o}nig, ``Electroweak measurements on the Z'', talk at DESY-ECFA 
LC Workshop, March 20-23, 1999, Oxford, U.K., to appear in the proceedings.

\bibitem{DESY99070}
D.~Bardin, P.~Christova, M.~Jack, L.~Kalinovskaya, A.~Olchevski, S.~Riemann,
  and T.~Riemann, ``{\tt ZFITTER} v.6.11 -- a semi-analytical program for
  fermion pair production in $e^+e^-$ annihilation'', DESY preprint 99-070
  (1999), hep-ph/9908433.

\bibitem{Beenakker:1991mb}
W.~Beenakker, F.~Berends, and S.~C. van~der Marck, {\em Nucl. Phys.} {\bf B349}
  (1991) 323.

\bibitem{Burgers:BHM}
G.~Burgers, W.~Hollik, and M.~Martinez, Fortran program {\tt BHM}.

\bibitem{Jadach:1994yv}
S.~Jadach, B.~F.~L. Ward, and Z.~Was, {\em Comput. Phys. Commun.} {\bf 79}
  (1994) 503--522.

\bibitem{Jadach:1999kz}
S.~Jadach, B.~F.~L. Ward, and Z.~Was, ``The Monte Carlo program {\tt KORALZ} for
the lepton or quark pair production at LEP / SLC energies: from version 4.0 
to version 4.04'', preprint CERN-TH. 99-119 (May 1999),
submitted to Comput. Phys. Commun., hep-ph/9905205.  

\bibitem{Montagna:1998kp}
G.~Montagna, O.~Nicrosini, F.~Piccinini, and G.~Passarino, {\em Comput. Phys.
  Commun.} {\bf 117} (1999) 278.

\bibitem{Bardin:1989di}
D.~Bardin, M.~S. Bilenky, G.~Mitselmakher, T.~Riemann, and M.~Sachwitz, {\em Z.
  Phys.} {\bf C44} (1989) 493.

\bibitem{Bardin:1989cw}
D.~Bardin, M.~Bilenky, A.~Chizhov, A.~Sazonov, Y.~Sedykh, T.~Riemann, and
  M.~Sachwitz,\\ {\em Phys. Lett.} {\bf B229} (1989) 405.

\bibitem{Bardin:1991de}
D.~Bardin, M.~Bilenky, A.~Sazonov, Y.~Sedykh, T.~Riemann, and M.~Sachwitz, {\em
  Phys. Lett.} {\bf B255} (1991) 290--296.

\bibitem{Bardin:1991fu}
D.~Bardin, M.~Bilenky, A.~Chizhov, A.~Sazonov, O.~Fedorenko, T.~Riemann, and
  M.~Sachwitz, {\em Nucl. Phys.} {\bf B351} (1991) 1--48.

\bibitem{Christova:1999cc}
P.~C. Christova, M.~Jack, and T.~Riemann, {\em Phys. Lett.} {\bf B456} (1999)
  264.

\bibitem{DESY99037}
P.~Christova, M.~Jack, S.~Riemann, and T.~Riemann, ``Predictions of {\tt ZFITTER}
  v.6 for fermion-pair production with acollinearity cut'', DESY preprint 99-037
  (1999), hep-ph/9908289.

\bibitem{Passarino:1982zp}
G.~Passarino, {\em Nucl. Phys.} {\bf B204} (1982) 237--266.

\bibitem{Bardin:1992jc2}
D.~Bardin {\em et~al.}, ``{\tt ZFITTER} v.4.5: An analytical program for
  fermion pair production in $e^+e^-$ annihilation'', preprint CERN-TH. 6443/92
  (May 1992), hep-ph/9412201.

\bibitem{SITGES99}
M.~Jack, T.~Riemann, ``Prospects and Problems in Fermion-Pair Production'',
to appear in the proceedings of the ECFA-DESY LC Workshop,  
April 28 - May 5, 1999, Sitges (Barcelona), Spain, hep-ph/9909401.

\bibitem{Bardin:1995aa}
D.~Bardin, G.~Passarino, and {W. Hollik (eds.)}, ``Reports of the working group
  on precision calculations for the {$Z$} resonance'', report CERN 95--03
  (1995), hep-ph/9709229.

\bibitem{Bardin:1999gt}
D.~Bardin, M.~Gr{\"u}newald, and G.~Passarino, ``Precision calculation project
  report'', {\tt hep-ph/9902452}.

\bibitem{Riemann:199200}
S.~Riemann, unpublished comparisons (1992).

\bibitem{Montagna:1993mf}
G.~Montagna, O.~Nicrosini, and G.~Passarino, {\em Phys. Lett.} {\bf B309}
  (1993) 436--442.

\bibitem{Bilenkii:1989zg}
M.~Bilenky and A.~Sazonov, ``{QED} corrections at {$Z^0$} pole with realistic
  kinematical cuts'', Dubna preprint JINR-E2-89-792 (1989).

\bibitem{MBilenky:1989ab}
M.~Bilenky and A.~Sazonov, {\tt ZFITTER} Fortran routines for photonic
  corrections with acollinearity and acceptance cuts, unpublished (1989).

\bibitem{zfitter:v6.0406m}
D.~Bardin, P.~Christova, M.~Jack, L.~Kalinovskaya, A.~Olshevski, S.~Riemann,
  and T.~Riemann, Fortran program {{\tt ZFITTER}} v.6.04 (April 21, 1999),
  v.6.06 (May 13, 1999).

\bibitem{DESY:1999??}
P.~Christova, M.~Jack, and T.~Riemann, extended write-up in
preparation.

\bibitem{zfitter:v5.2021m}
D.~Bardin, P.~Christova, L.~Kalinovskaya, A.~Olshevski, and S.~Riemann, Fortran
  program {{\tt ZFITTER}} v.5.20 (Feb 17, 1999), v.5.21 (March 09, 1999).

\bibitem{Passarino:199800}
G.~Passarino, private communication, November 1998.

\bibitem{Jadach:1998jb}
S.~Jadach, B.~F.~L. Ward, and Z.~Was, {\em Phys. Lett.} {\bf B449} (1999)
  97--108.

\bibitem{Jadach:1999gz}
S.~Jadach, B.~Pietrzyk, E.~Tournefier, B.~Ward, and Z.~Was,
  ``Initial-final-state interference in the {Z} line-shape'', preprint CERN-TH.
  99-217 (Jul 1999), {\tt hep-ph/9907547}.
\end{thebibliography}
\endgroup
\end{document}